\documentclass{article}
    
    \usepackage{lineno,hyperref}
    \modulolinenumbers[5]
    
\usepackage{natbib}
    \usepackage{graphicx}
    \usepackage{amsmath}
    \usepackage{amssymb}
    \usepackage{url}
    \usepackage{xcolor}
    \usepackage{comment}
    
\begin{document}
    
    \title{Dual Labor Market \\ and the ``Phillips Curve Puzzle"}
        \date{\small \today}
\author{Hideaki Aoyama$^{1,2}$, Corrado Di Guilmi$^{3,4,5}$,
Yoshi Fujiwara$^6$,\\[5pt]
and Hiroshi Yoshikawa$^{2,7}$\\[10pt]
\small $^1$RIKEN iTHEMS, Wako, Saitama 351-0198, Japan\\
\small $^2$Research Institute of Economy, Trade and Industry\\
\small (RIETI), Tokyo 100-0013, Japan\\
\small $^3$Economics Discipline Group,
\small University of Technology Sydney, Australia,\\
\small $^4$Centre for Applied Macroeconomic Analysis, \\
\small Australian National University, Canberra, Australia\\
\small $^5$Center for Computational Social Science,\\
\small Kobe University, Kobe 657-8501, Japan\\
\small $^6$University of Hyogo, Kobe 650-0047, Japan\\
\small $^7$Rissho University, Tokyo 141-8602, Japan}
   
\maketitle

\begin{abstract}
Low inflation was once a welcome to both policy makers and the public. However, Japan's experience during the 1990's changed the consensus view on price of economists and central banks around the world. Facing deflation and zero interest bound at the same time, Bank of Japan had difficulty in conducting effective monetary policy. It made Japan's stagnation unusually prolonged. 
Too low inflation which annoys central banks today is translated into the ``Phillips curve puzzle''. In the US and Japan, in the course of recovery from the Great Recession after the 2008 global financial crisis, the unemployment rate had steadily declined to the level which was commonly regarded as lower than the natural rate or NAIRU. And yet, inflation stayed low.
In this paper, we consider a minimal model of dual labor market to explore what kind of change in the economy makes the Phillips curve flat. The level of bargaining power of workers, the elasticity of the supply of labor to wage in the secondary market, and the composition of the workforce are the main factors in explaining the flattening of the Phillips curve. We argue that the changes we consider in the model, in fact, has plausibly made the Phillips curve flat in recent years.
\end{abstract}
    

    
\newcommand{\wtot}{W}
\newcommand{\Lone}{L_1}
\newcommand{\Ltwo}{L_2}
\newcommand{\Lonej}{L_{1,j}}
\newcommand{\Ltwoj}{L_{2,j}}
\newcommand{\wone}{w_1}
\newcommand{\wonej}{w_{1,j}}
\newcommand\wtwo{w_2}
\newcommand{\barw}{\bar{w}}
\newcommand{\Ltwojdem}{L_{2,j}^{\rm (d)}}
\newcommand{\Ltwodem}{L_2^{\rm (d)}}
\newcommand{\Ltwosup}{L_2^{\rm (s)}}
\newcommand{\zfun}{Z(\alpha,c,\beta,\gamma,g,v)}
\newcommand{\bH}{\boldsymbol{\rm H}}
\newcommand{\bM}{\boldsymbol{\rm V}}
\newcommand{\bT}{\boldsymbol{\rm T}}


\section{Introduction}
    
Low inflation was once a welcome to both policy makers and the public. However, Japan's experience during the 1990's changed the consensus view on price of economists and central banks around the world; After the financial bubble burst at the beginning of the 1990's, Japan lapsed into deflation. During the course, the Bank of Japan (BOJ) kept cutting the nominal interest rate down to zero. Facing deflation and zero interest bound at the same time, BOJ had difficulty in conducting effective monetary policy. It made Japan's stagnation unusually prolonged.

The ``Japan problem'' made economists aware of long-forgotten danger of deflation. In the prewar period, deflation was a menace to the economy, and its danger was emphasized by famous economists such as 
\citet{keynes2010consequences}
and
\citet{fisher1933debt}.
To prevent deflation, central bank must seek low inflation rather than zero inflation or stable price level. Today, following this idea, many central banks including The US Federal Reserve, BOJ, and European Central Bank target at two percent inflation of consumer price index. However, few central banks have been successful in achieving this goal in any satisfactory way.

Too low inflation which annoys central banks today is translated into the “Phillips curve puzzle”. The benchmark Phillips curve is as follows
\citep{phillips1958relation,friedman1968role}:
\begin{equation}
\pi_t=a\,(u-u^*)+b\, \pi_t^*,
\label{eqn:first}
\end{equation}
where $\pi$ and $u$ are inflation and the unemployment rate, respectively. $\pi^*$ is either inflationary expectation or inertia of past inflation. $u^*$ is the natural rate of unemployment or the NAIRU (Non-Accelerating Inflation Rate of Unemployment). According to conventional macroeconomics, Eq.~(\ref{eqn:first}) or the Phillips curve determines inflation.

In the US and Japan, in the course of recovery from the Great Recession after the 2008 global financial crisis, the unemployment rate had steadily declined to the level which was commonly regarded as lower than NAIRU, $u^*$. And yet, inflation stayed low: 0.5\% for Japan and 1.5\% for the US. 

In terms of the Phillips curve, Eq.~(\ref{eqn:first}), two things have been pointed out.
First, coefficient $b$ for inflationary expectations or lagged inflation declined significantly almost to zero in recent years. 
\citet[Figure 7]{blanchard2018should},
for example, found that $b$ which was almost zero in the early 1960's, rose sharply to one in the late 1960's, had stayed there for thirty years, and then declined suddenly to zero at the beginning the 2000's.
While the decline in inflation from the '90s had been often attributed to better policy management, and even the tern ``Great Moderation" was coined \citep{Clarida00}, more recent analyses identify the anchoring of inflation expectations for the change in the trade-off between inflation and unemployment \citep{BARNICHON2020,Blanchard16}. 
\cite{greenspan2001transparency}
left the following remark: ``Price stability is best thought of as an environment in which inflation is so low and stable over time that it does not materially enter into the decisions of households and firms.''

The second factor is a change of coefficient $a$ for the unemployment rate in Eq.~(\ref{eqn:first}). A decline of $a$ entails lower inflation than otherwise when the unemployment rate declines. Some researchers even argue that unemployment no longer has an effect on inflation, at least over some unemployment and inflation range. 

\begin{figure}
\centering
(a)\\
\includegraphics[width=0.7\textwidth]{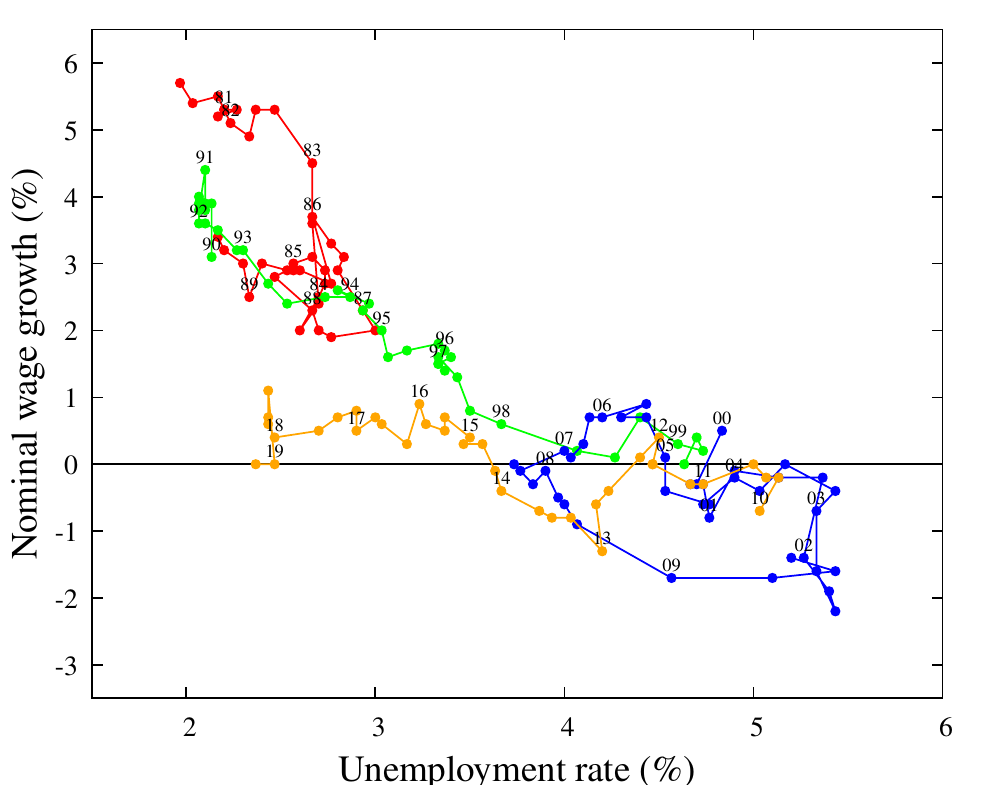}\\
(b)\\
  \includegraphics[width=0.7\textwidth]{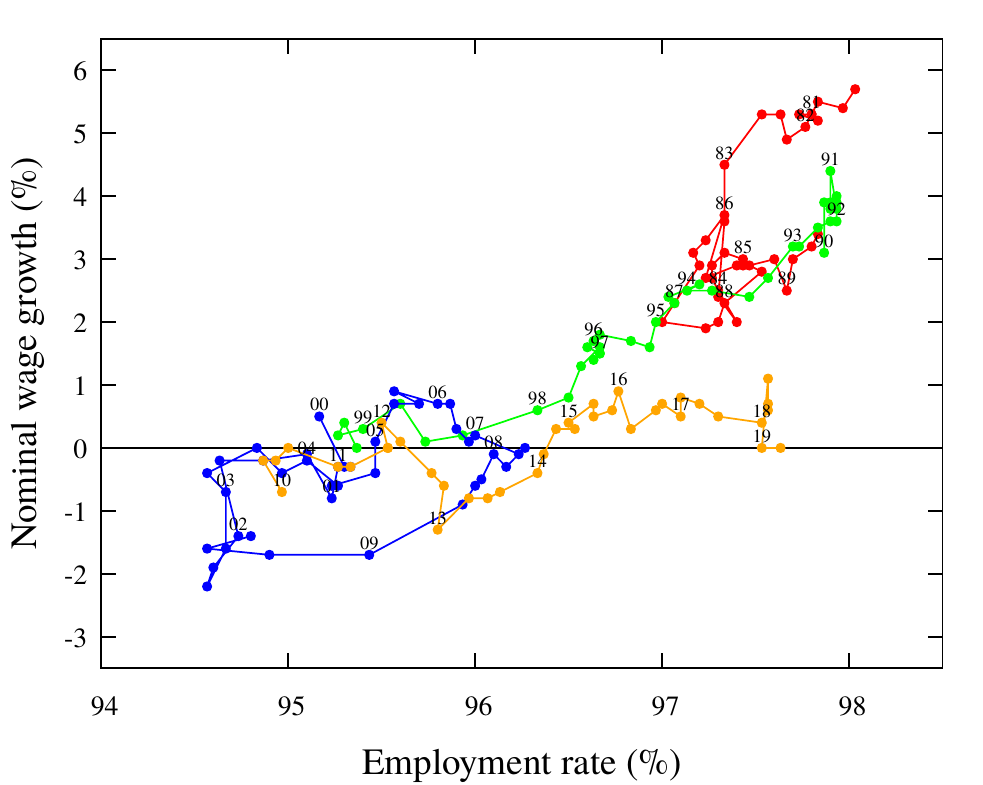}
\caption{\textbf{Japan's Phillips curve:
         Unemployment rate and nominal wage growth of 157 quarters from 1980.$\Pi$ to 2019.$\Pi$.}
         (a) The abscissa is the unemployment rate in \%.
         (b) The abscissa is reversed and is the employment rate (=100-unemployment rate (\%)).
         Colors: 1980’s (red), 1990’s (green), 2000’s (blue), 2010’s (orange).
        The last several years' data (in yellow) show definite deviation from those of the earlier years, with nominal wage growth is low in spite of the low unemployment.
        Sources are \cite{kinro} and \cite{rodo}.}
\label{fig:PC}
\end{figure}

Figure \ref{fig:PC} (a) displays Japan's Phillips curve, namely, the quarterly relation between the unemployment rate and nominal wage growth for 1980.II-2019.II. We can indeed observe that the Phillips curve has flattened in recent years.

These are the ``Phillips curve puzzle''. 
In this paper, we leave the first problem untouched: we simply take $b$ as zero, and focus on the second problem, namely why $a$ gets smaller in Eq.~(\ref{eqn:first}). For this purpose, we consider a minimal model of dual labor market \citep{MCDonaldSolow81,MCDonaldSolow85, Gordon2017new} to explore what kind of change in the economy makes the Phillips curve flat. 
In the model, the level of bargaining power of workers, the elasticity of the supply of labor to wage in the secondary market, and the composition of the workforce are the main factors in explaining the flattening of the Philips curve. 

Our main contribution is to provide a compact model to jointly consider factors that have been so far mostly investigated separately by the literature. Our main finding is that the change in shape in the relationship between the level of economic activity and inflation can only be explained by the joint effects of these four factors. The structural modifications in the labor market that occurred in Japan over the last three decades have determined the dramatic shift in the relationship between inflation and level of economic activity.

The remainder of the paper is organized as follows. Section \ref{sec:model} introduces the theoretical model. Sections \ref{sec:analytical} and \ref{sec:numerical} presents the analytical and numerical results, respectively. 
Finally, section \ref{sec:discussion} provides some concluding remarks.
 
\section{The Model}\label{sec:model}
The model assumes a dual labor market consisting of primary labor and secondary workers \citep{MCDonaldSolow81,MCDonaldSolow85,Gordon2017new}. 
As in\linebreak
\cite{DiGuilmiFujiwara20}, we identify as secondary workers all the employees without a permanent contract (agency, temporary, and part-time). 

Firms are heterogeneous in size and efficiency, but adopt the same production function.
Each firm produces a homogeneous goods by employing only labor, composed by primary and secondary workers.
Specifically, the firm $j$ $(j=1,\cdots,N$) employs $\Lonej$ primary workers with wage  $\wonej$ and $\Ltwoj$ secondary workers with wage $\wtwo$.
There are $\Lone=\sum_{j=1}^N \Lonej$ primary workers and $\Ltwo=\sum_{j=1}^N \Ltwoj$ secondary workers employed.

Primary workers are a fixed endowment for each firm.\footnote{The Japanese firm rarely lays off its primary workers.
In 2020, for example, in the mid of Covid--19 recession, real GDP fell by unprecedented 28.1\%,
and yet, the unemployment rate rose slightly only to 2.8\%.}
Accordingly, $L_{1,j}$ is a given constant.
In contrast, firm freely changes the level of secondary workers.
Following the empirical findings of \cite{Munakata16}, the wage of secondary workers is determined by the market and is uniform across firms.

The output of firm $j$ is determined as follows:
    \begin{equation}
        Y_{j}=A_{j} (\Lonej+c \, \Ltwoj)^{\alpha},
        \label{eq:output}
    \end{equation}
where $\alpha \in (0,1)$ . 
$A_j$ is firm-specific total factor productivity (TFP). 
The parameter $c\in (0,1)$ quantifies the productivity of secondary labor relative to that of primary labor.

The profit is given by
    \begin{equation}
        \Pi_{j} = Y_{j}-\Lonej \wonej-\Ltwoj \wtwo.
        \label{eq:profit}
    \end{equation}
In order to mimic the firm-level bargaining process that is prevailing in Japan, the primary workers' wage is set in a two-step process for each employer. Assuming a fixed endowment of primary workers (or \emph{insiders}) $\Lonej$ for each firm, in the first stage firm and primary workers determine the number of secondary workers (or \emph{outsiders}) $\Ltwoj$ to be hired. Assuming a perfectly competitive market for secondary workers, firms take the secondary wage $\wtwo$ as given. Once the number of secondary workers is determined, firms and insiders share the surplus defined by revenue less the wages paid to secondary workers through a Nash bargaining \citep{SM}.

\subsection*{First stage maximization: profit}
Firm maximizes its profit (\ref{eq:profit}) by choosing the number of secondary workers or outsiders:
    \begin{equation}
        \max_{\Ltwoj} \left[ \Pi_j \right].
        \label{eq:maximize1}
    \end{equation}
This determines demand for secondary workers $\Ltwojdem$ of firm $j$ as follows:
    \begin{equation}
        \Ltwojdem=\frac{1}{c}\left[
        -\Lonej+\left(\frac{\alpha c A_j}\wtwo\right)^{1/(1-\alpha)}
        \right].
        \label{eq:max1res}
    \end{equation}
The total demand for secondary workers in the economy as a whole is then:
    \begin{equation}
      \Ltwodem =\frac{1}{c}
       \left[-\Lone+
        \bigg(\frac{\alpha\,c\, A}\wtwo\bigg)^{1/(1-\alpha)}
       \right], \label{eq:demL2}
       \end{equation}
where $A$ is the following nonlinear sum of $A_j$:
       \begin{equation}
       A=\left(\sum_{j=1}^N A_j^{1/(1-\alpha)}\right)^{(1-\alpha)}.   
       \label{eq:defA}   
    \end{equation}
    
The level of output of firm $j$ is 
     \begin{equation}
        Y_j=A_j
        \bigg(\frac{\alpha c A_j}\wtwo\bigg)^{\alpha/(1-\alpha)}.
        \label{eq:yjres}
    \end{equation}
    
\begin{figure}[t]
        \centering
        \includegraphics[width=0.8\textwidth]{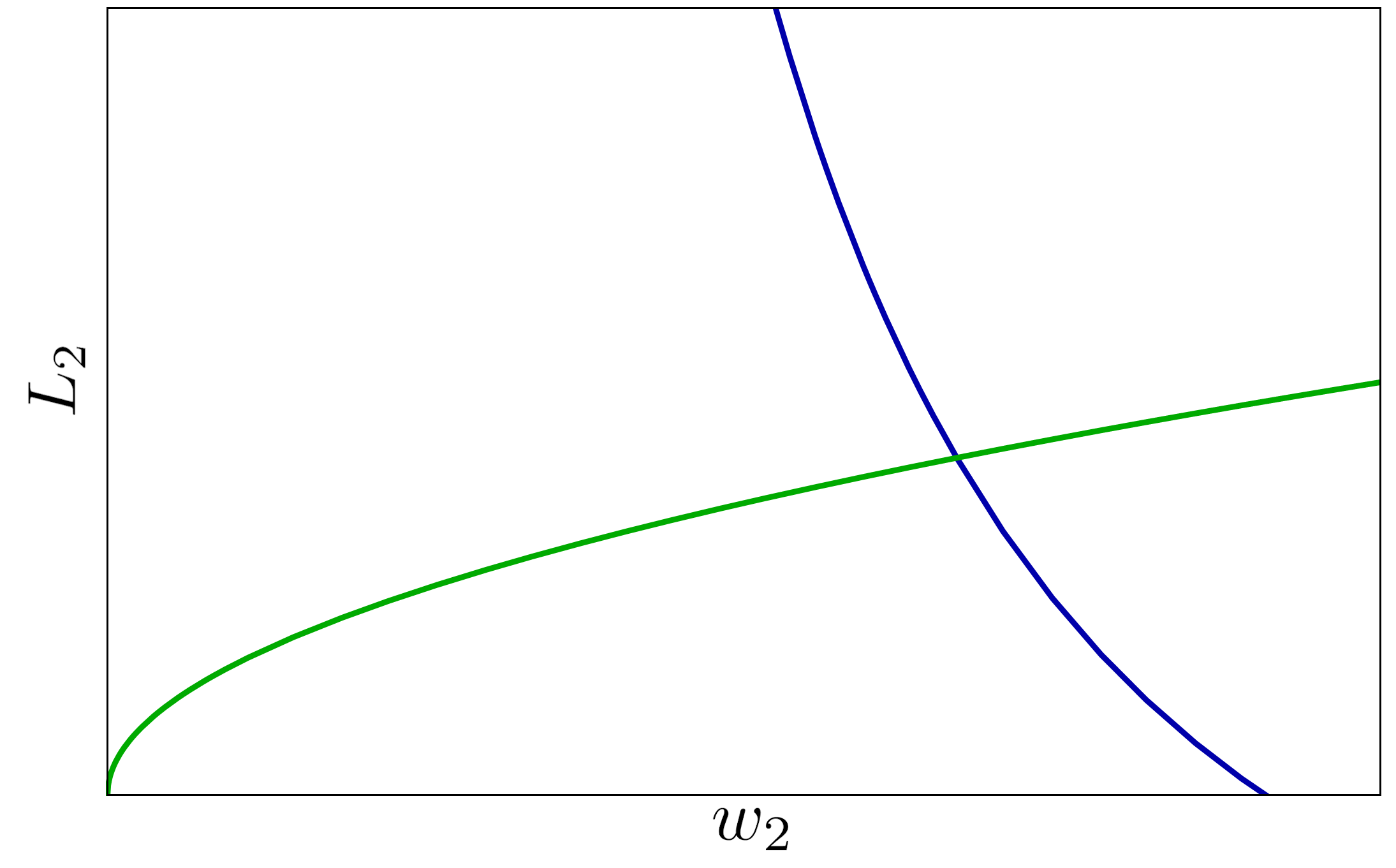}
        \caption{\textbf{Demand function $\Ltwodem$ (Eq.\eqref{eq:demL2}) in blue and supply function $\Ltwosup$ (Eq.\eqref{eq:supply}) in green.}}
        \label{fig:SupplyDemand}
    \end{figure}

We assume the following supply function of secondary workers $\Ltwo$:
    \begin{equation}
        \Ltwosup=B\, \wtwo^\beta.
        \label{eq:supply}
    \end{equation}
    where $\beta$ is the Frisch elasticity.
Matching demand for and supply of secondary workers, $\Ltwosup=\Ltwodem$,
we obtain the following nonlinear equation for $\wtwo$:
    \begin{equation}
       B\, \wtwo^\beta=\frac{1}{c}
       \left[-\Lone+
        \bigg(\frac{\alpha\,c\, A}\wtwo\bigg)^{1/(1-\alpha)}
       \right], \label{eq:detw2}
       \end{equation}
Demand for and supply of secondary labor as functions of $w_2$ are shown on $(\Ltwo, \wtwo)$ plane in Fig.~\ref{fig:SupplyDemand}. 
It can be shown that the solution of Eq.~(\ref{eq:detw2}) always exists.
    
\subsection*{Second stage maximization: primary workers' wage}
Firm and primary workers (insiders) determine the wage of primary workers $w_{1,j}$ through a Nash bargaining according to:
    \begin{equation}
       \max_{\wonej} \left[(\Lonej \wonej)^\gamma (\Pi_{j})^{(1-\gamma)}\right],
        \label{eq:maximize2}
    \end{equation}
where $\gamma\in (0,1)$ indicates bargaining power of primary workers.

The Nash bargaining determines $\wonej$ as follows:
    \begin{equation}
        \wonej=\gamma\frac{
        Y_j -\Ltwoj \wtwo}{\Lonej}.
        \label{eq:max2resw1j}
    \end{equation}

By combining Eqs.(\ref{eq:max1res}) and (\ref{eq:max2resw1j}), we find that
\begin{equation}
    \Pi_j=\frac{1-\gamma}{\gamma}\Lonej\wonej.
    \label{eq:profitsol}
\end{equation}

In the following, we study the relationship between the total employment of workers, 
    \begin{equation}
        L=\Lone+\Ltwo=\Lone+B\wtwo^\beta,
    \label{eq:defL}
    \end{equation}
and the average wage,
    \begin{equation}
    \barw=\frac{\wtot}{L}.
    \label{eqn:wbardef}
    \end{equation}
In Eq.\eqref{eqn:wbardef}, $W$ is the total earnings of all the workers:
    \begin{align}
        \wtot&=\sum_{j=1}^{N}\left(\Lonej \wonej\right)+\Ltwo\wtwo\nonumber\\
        &=\gamma\sum_{j=1}^N Y_j + (1-\gamma)\Ltwo\wtwo\nonumber\\
        &=\gamma A \left(\frac{\alpha cA}{\wtwo}\right)^{\alpha/(1-\alpha)}
        +(1-\gamma)B\wtwo^{1+\beta}.
    \label{eq:defW}
    \end{align}
We use the result of the second maximization, Eq.\eqref{eq:max2resw1j}.
Because $L$ and $\barw$ are functions of $A$, we obtain functional relationship between $L$ and $\barw$ by eliminating $A$, while keeping other parameters \{$L_1, c,\ \alpha,$ $\ B,\ \beta,$ $\gamma$\} fixed.
The curve $w(L)$ is our Phillips curve, which models the relationship shown in Fig.\ref{fig:PC}(b).

Our Phillips curve is expressed in level of wage (rather than wage inflation) since, as we show below, the rate of change in wage is implied by the wage level.
 
The parameters and variables of this model are listed in Table \ref{tab:v}.
The model is extremely parsimonious, with only seven free parameters: $A, L_1, c, \alpha, B, \beta, \gamma$.
However, because of nonlinearities, its solution is not trivial and able to generate a set of interesting results.

\clearpage
 \begin{table}[t]
 \begin{tabular}{cl}
 \hline
       \multicolumn{2}{l}{Output}\\
      $A$ &  Nonlinear sum of the total factor productivity $A_j$ (Eq.(\ref{eq:defA}))\\
      $\Lone$& Total number of primary workers\\
      $L_2$& Total number of secondary workers employed (Eq.(\ref{eq:max1res}))\\
      $c$& Secondary workers' productivity coefficient\\
      $\alpha$& Output exponent  \\
      \hline
      \multicolumn{2}{l}{Supply of secondary workers}\\
      $B$& Coefficient of labor supply for secondary workers\\
      $\beta$& Elasticity to wage for secondary workers\\
      \hline
       \multicolumn{2}{l}{Nash Bargaining}\\
      $w_{1,j}$& The wage of the primary workers at firm $j$ (Eq.~(\ref{eq:max2resw1j}))\\
      $w_2$& The wage of secondary workers (Eq.~(\ref{eq:detw2}))\\
      $\gamma$& Bargaining power of primary workers\\
      \hline
       \end{tabular}
 \caption{\textbf{List of the parameters and variables of the model.}
 Variables are determined by the equation referred in parentheses.}
 \label{tab:v}
 \end{table}

\clearpage

\section{Solving the Model}\label{sec:analytical}
Toward the goal of solving the model, we first rewrite Eq.\eqref{eq:detw2} as follows: 
 \begin{equation}
     \frac{\wtwo}{\alpha\, c\, A}=\left(\Lone + c\,B\,\wtwo^\beta\right)^{-(1-\alpha)}.
 \label{eq:detw2b}
 \end{equation}
By introducing the following scaled variable $v$
 \begin{equation}
     v\equiv\frac{\ \Lone^{\beta(1-\alpha)}}{\ (\alpha\, c \,A)^\beta\ }\wtwo^\beta.
     \label{eq:vdef}
 \end{equation}
we can rewrite Eq.\eqref{eq:detw2b} as follows:
 \begin{equation}
     v=\left(1+g\, v\right)^{-\beta(1-\alpha)},\label{eq:veq}
 \end{equation}
 where
 \begin{equation}
  g\equiv c\,B\left(\alpha\, c\, A\right)^\beta \Lone^{-1-\beta(1-\alpha)}.\label{eq:gdef}
\end{equation}

Recall that solving the model amounts to finding the equilibrium $w_2$ which is equivalent to $v$.
Thus, we focus on Eq.\eqref{eq:veq}.
We note here that both $v$ and $g$ are dimensionless quantities in Eq.\eqref{eq:veq} (see Appendix A).
This makes the following analysis straightforward.

With variable $v$, Eqs.\eqref{eq:defL} and \eqref{eq:defW} are written as follows:
\begin{align}
     L&=\Lone\left[1+\frac{g}{\ c\ } v\right],
     \label{eq:zL}\\[5pt]
     W&= \frac{\ g^{1/\beta}\Lone^{1+1/\beta}}{\alpha c (cB)^{1/\beta}}
     \left[
     \gamma\, v^{-\alpha/(\beta(1-\alpha))}+(1-\gamma)\,\alpha\, g\, v^{1+1/\beta}
     \right].
     \label{eq:zW}
\end{align}
This leads to the following average wage:
\begin{equation}
     \bar{w}=\frac{W}{L}=\left(\frac{L_1}{B}\right)^{1/\beta}\zfun.
     \label{eq:wbarg}
\end{equation}
The coefficient $(L_1/B)^{1/\beta}$ is the {\it only} factor with the same dimension as $\bar{w}$.
The function $\zfun$ is the following dimensionless function of the dimensionless parameters  $\alpha,c,\beta,\gamma,g$ and $v=v(g)$:   
\begin{align}
     &\zfun=\nonumber\\
     &\quad\frac{g^{1/\beta}}{\alpha\, c^{1+1/\beta}}
   \left[
     \gamma\, v^{-\alpha/(\beta(1-\alpha))}+(1-\gamma)\,\alpha\, g\, v^{1+1/\beta}
     \right]
     \Bigg/
    \left[1+\frac{g}{\ c\ } v\right].
    \label{eq:wbarZ}
 \end{align}
The Phillips curve  defined as the relationship between the average wage and employment,  $\bar{w}(L)$, is obtained by eliminating $g$ (and $v=v(g)$) from Eq.~(\ref{eq:zL}) and Eq.~(\ref{eq:wbarg}).
 
Now, the second term in the parentheses of the right-hand-side of Eq.\eqref{eq:veq}, $g v$ is the ratio $cL_2/L_1$. 
Therefore, if $\Lone$ and $\Ltwo$ measured in efficiency unit are different in order, we may approximate it around the larger term. In other words, if $\Lone\gg c \Ltwo$, namely, if the primary workers dominate in efficiency in production, we expand the right hand side around for $gv\ll 1$
or equivalently $g\ll 1$ (note that large $\Lone$ implies small $g$ because of Eq.\eqref{eq:gdef}).


The small-$g$ perturbative solution of Eq.\eqref{eq:veq} is the following:
\begin{equation}
v=1-\sigma g +\frac12 \sigma(1+3\sigma)g^2
          +\cdots,
          \label{eq:zl2}
\end{equation}
where $\sigma\equiv \beta(1-\alpha)$.
By substituting Eq.\eqref{eq:zl2} into Eqs.\eqref{eq:zL}, \eqref{eq:wbarg}, and \eqref{eq:wbarZ}, we obtain the followings:
\begin{align}
 L&=\Lone\left[1+\frac{g}{\, c \,}
    -\sigma\frac{g^2}{c}  
    +\cdots\right]
    \label{eq:Lres}\\
 \barw=& \left(\frac{L_1}{B}\right)^{1/\beta}
     \frac{\ g^{1/\beta}}{\alpha\, c^{1+1/\beta}}
     \left[\gamma+\left(\alpha  - \frac{\gamma}{c}\right)g +\cdots\right].
    \label{eq:resbarw}
\end{align}
In order to eliminate $g$ from these two equations and obtain a relationship between $L$ and $\barw$,
we first solve Eq.\eqref{eq:Lres} for small $g$:
\begin{equation}
    g=c\, \left(\frac{L}{\Lone}-1\right)
    +c^2 \sigma\left(\frac{L}{\Lone}-1\right)^2
    +\cdots.
\end{equation}
Note that this is a perturbative series in $L/ \Lone-1 =\Ltwo/ \Lone\ll 1$. 
By substituting this expression into Eq.\eqref{eq:resbarw},
we obtain the following expression of $\barw$:
 \begin{equation}
     \barw=\frac{\gamma}{\alpha\, c}\left(\frac{L_1}{B}\right)^{1/\beta}\left(\frac{L}{\Lone}-1\right)^{1/\beta}
     \left[1 +\frac{c(\alpha +\gamma-\alpha\gamma)-\gamma}{\gamma} \left(\frac{L}{L_1}-1\right)
     +\cdots \right].
     \label{eq:wbarLl}
 \end{equation}
 
Thus, the average wage $\barw$ is a monotonically increasing function of total employment, $L$. 
Namely, the Phillips curve has the expected sign of slope: it is
upward sloping on employment--wage plane, therefore, is
downward sloping on wage--unemployment plane.
We find from the leading term of this expression that the slope of the curve depends on two key factors, $\gamma/(c\, \alpha  B^{1/\beta})$ and $1/\beta$. 

The above analysis assumes $L_1 \gg c L_2$.
We can make similar analysis in the case of $L_1 \ll c L_2$.
It is given in Appendix B.

\section{Comparative Statics}\label{sec:numerical}
\begin{figure}[t]
    \centering
    \includegraphics[width=0.8\textwidth]{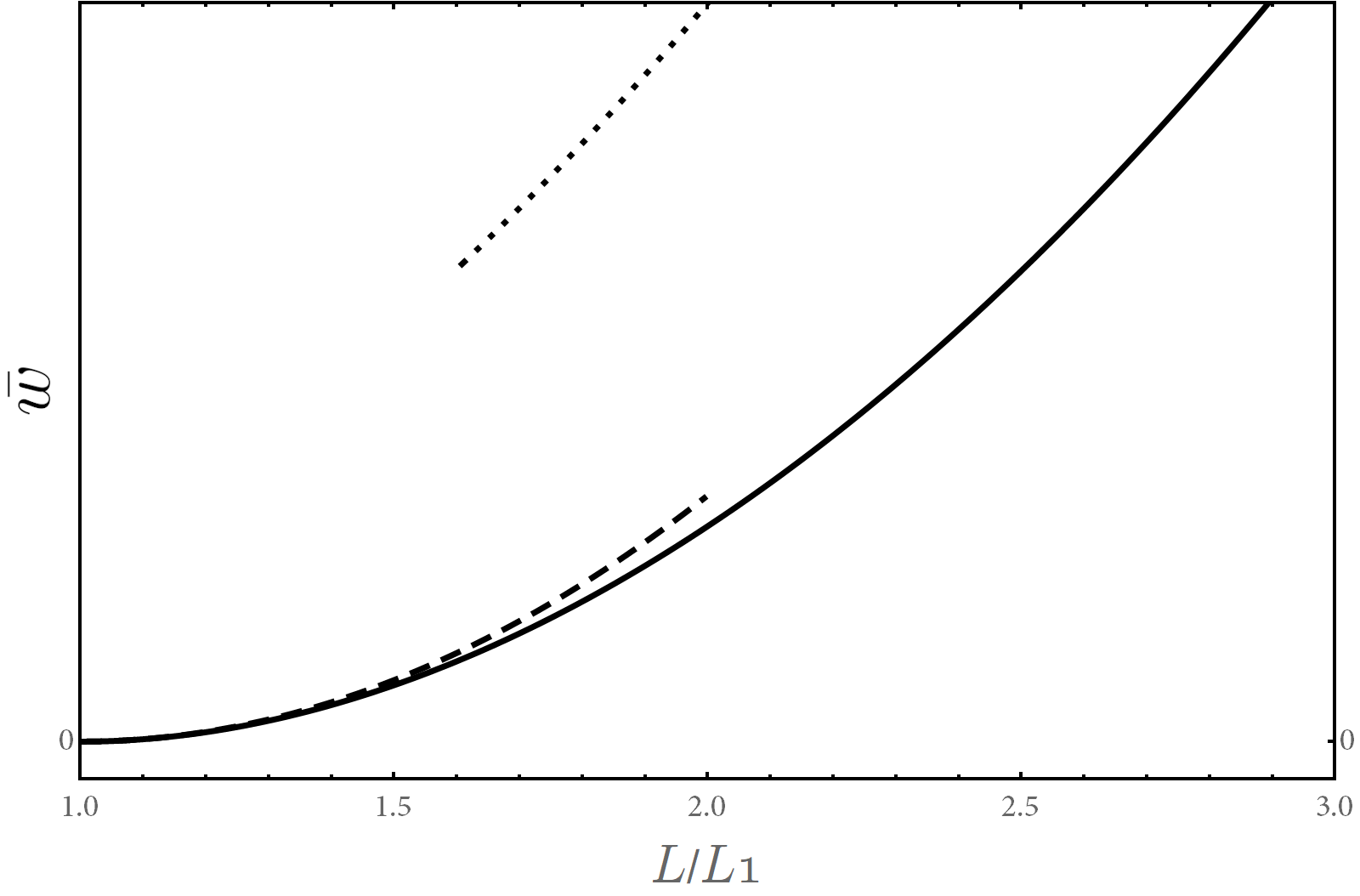}
    \caption{\textbf{The exact and approximate Phillips curve.}
    The solid curve: the exact solution $\bar{w}(L)$ (Eqs.~\eqref{eq:veq}, \eqref{eq:zL}, \eqref{eq:wbarg}),
    the dashed curve: the analytical solution in case primary workers dominate (the leading term of Eq.\eqref{eq:wbarLl}),
    and
    the dotted curve: the analytical solution in case secondary workers dominate (the leading term of Eq.\eqref{eq:wbarLs}).
    The last approximation is not valid in this range of $L/L_1$.
   The parameters are chosen to be $\alpha=c=\beta=\gamma=0.5$.}
    \label{fig:Phillips1}
\end{figure}

In order to estimate the effects of the parameter, we provide here a numerical comparative statics exercise. For the numerical calculations, the parameters are estimated as follows: $c=0.5$ \citep[following][]{DiGuilmiFujiwara20}; $\gamma=0.5$ \citep{Carluccio15}; $\beta \in [0.7,0.1]$ \citep{Kuroda07}, while other parameters are calibrated.
The Phillips curve is plotted in Fig.~\ref{fig:Phillips1}, 
where the solid curve is the numerical solution of Eqs.~\eqref{eq:veq}, \eqref{eq:zL}, \eqref{eq:wbarg} and the dashed and dotted curves are  the analytical solutions Eq.\eqref{eq:wbarLl} and Eq.\eqref{eq:wbarLs}, respectively.
We find that 
 the analytical solution \eqref{eq:wbarLl} for the case when when primary workers are dominant provides a reliable approximation for it mimics well the original function
 in this range of $L/L_1$.

\begin{figure}[t]
    \centering
    \includegraphics[width=0.8\textwidth]{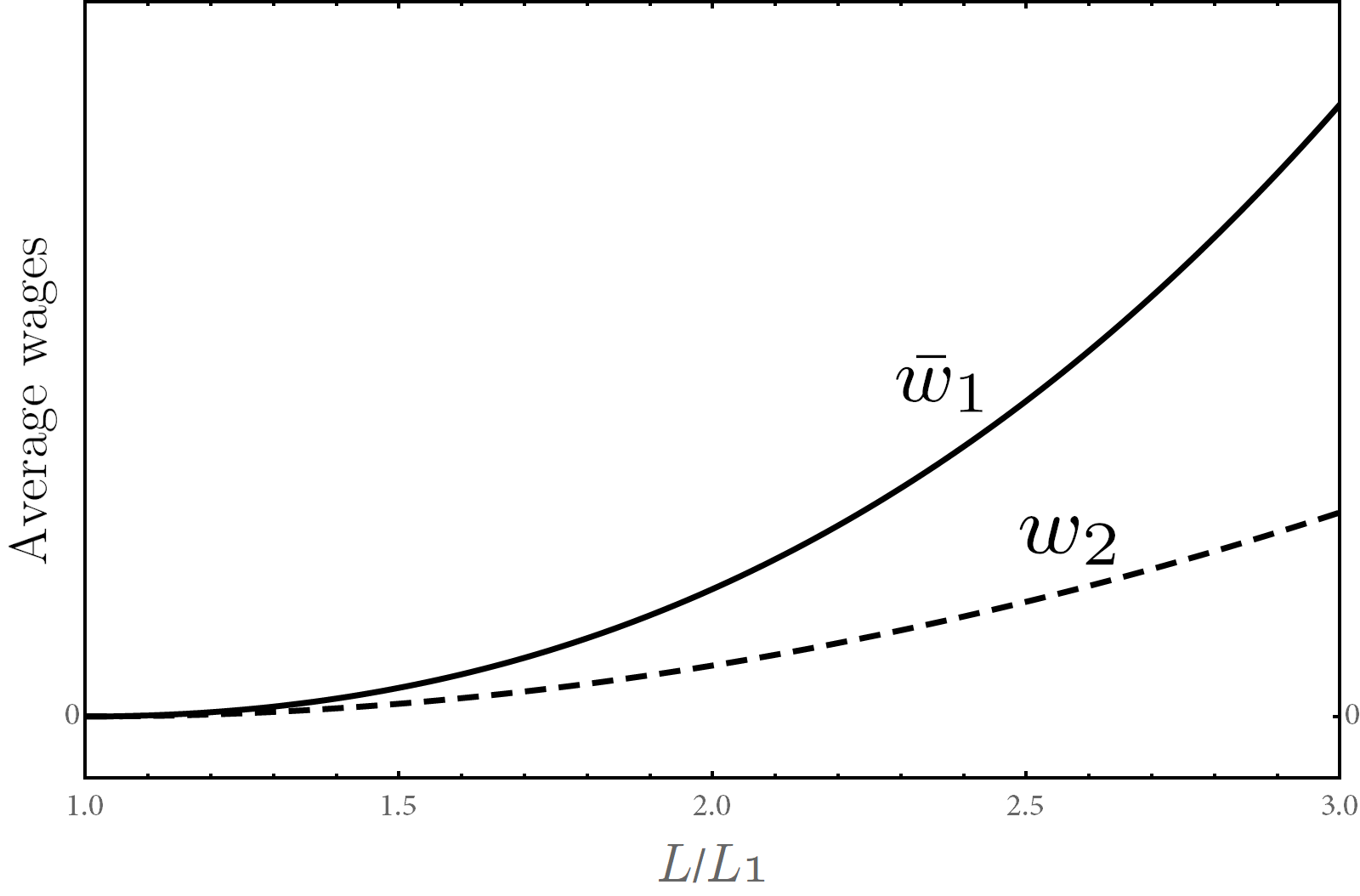}
    \caption{\textbf{The behavior of the wages $\bar{w}_1$ (solid curve) and $w_2$ (dashed curve).}
    The parameters are chosen to be the same with Fig.~\ref{fig:Phillips1}}
    \label{fig:w12}
\end{figure}

The wages of primary and secondary workers, 
\begin{equation}
    \bar{w}_1\equiv \frac1{L_1}\sum_{j=1}^N L_{1,j} w_{1,j}
\end{equation}
and $w_2$ as functions of total employment $L$ are shown in Fig.~\ref{fig:w12}.
It is interesting to observe that the wages of primary workers determined by Nash bargaining increase more than market--determined wages of secondary workers when the total employment increases.

Our goal is to find the answer for the question why the Phillips curve flattened in recent years.
For this purpose, we explore how the slope of Phillips curve depends on parameters $c, B, \gamma$,  and $\beta$.

\begin{figure}
\centering
  \includegraphics[width=0.8\textwidth]{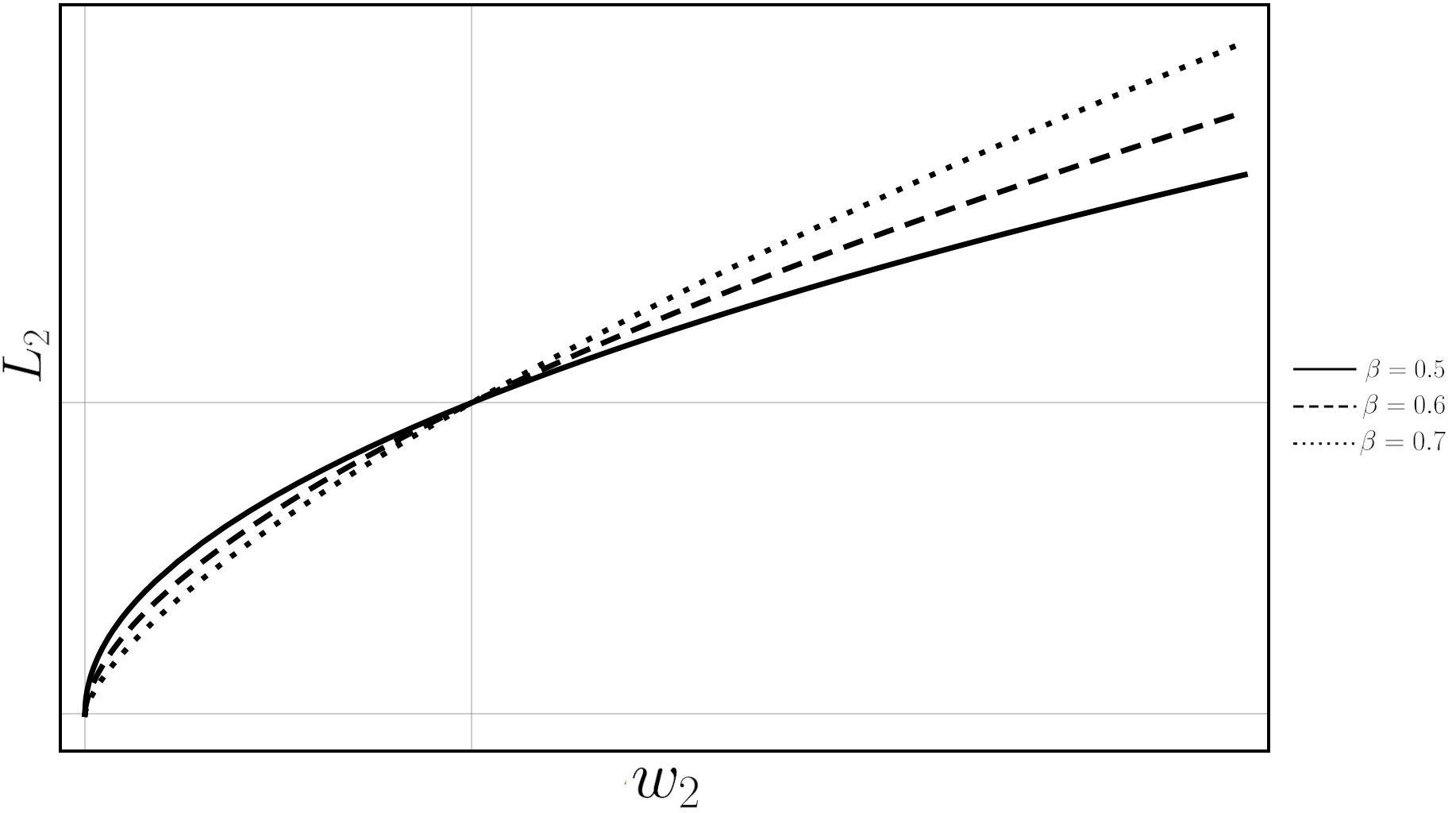}
\caption{\textbf{The meaning of the parametrizaion of the supply function with varying $\beta$.}}
\label{fig:beta}
\end{figure}

Variations of parameter $\beta$, however, require careful consideration:
in the supply function Eq.\eqref{eq:supply}, the coefficient $B$ has a dimension that depends on $\beta$.
Therefore, when changing the value of $\beta$, keeping the same numerical value of $B$ does not make sense. 
One way of making clear the dimensional consideration is to set $B=L_1/w_0^\beta$ so that
\begin{equation}
    \Ltwosup=L_1\, \left(\frac{w_2}{w_0}\right)^\beta.
\end{equation}
Accordingly, the supply function is parametrized by $w_0$ and $\beta$ instead of $B$ and $\beta$. 
In this parametrization, Eq.\eqref{eq:wbarg} is written as follows:
\begin{equation}
     \bar{w}=w_0\, \zfun.
     \label{eq:wbar2g}
\end{equation}
which makes the dimensionality trivial.
Using Eq. \eqref{eq:wbar2g}, when varying $\beta$, we keep $w_0$ constant and vary the value of $\beta$ in $\zfun$ in the above equation. This situation is illustrated in Fig.~\ref{fig:beta}.

\begin{figure}
    \centering
    \includegraphics[width=0.95\textwidth]{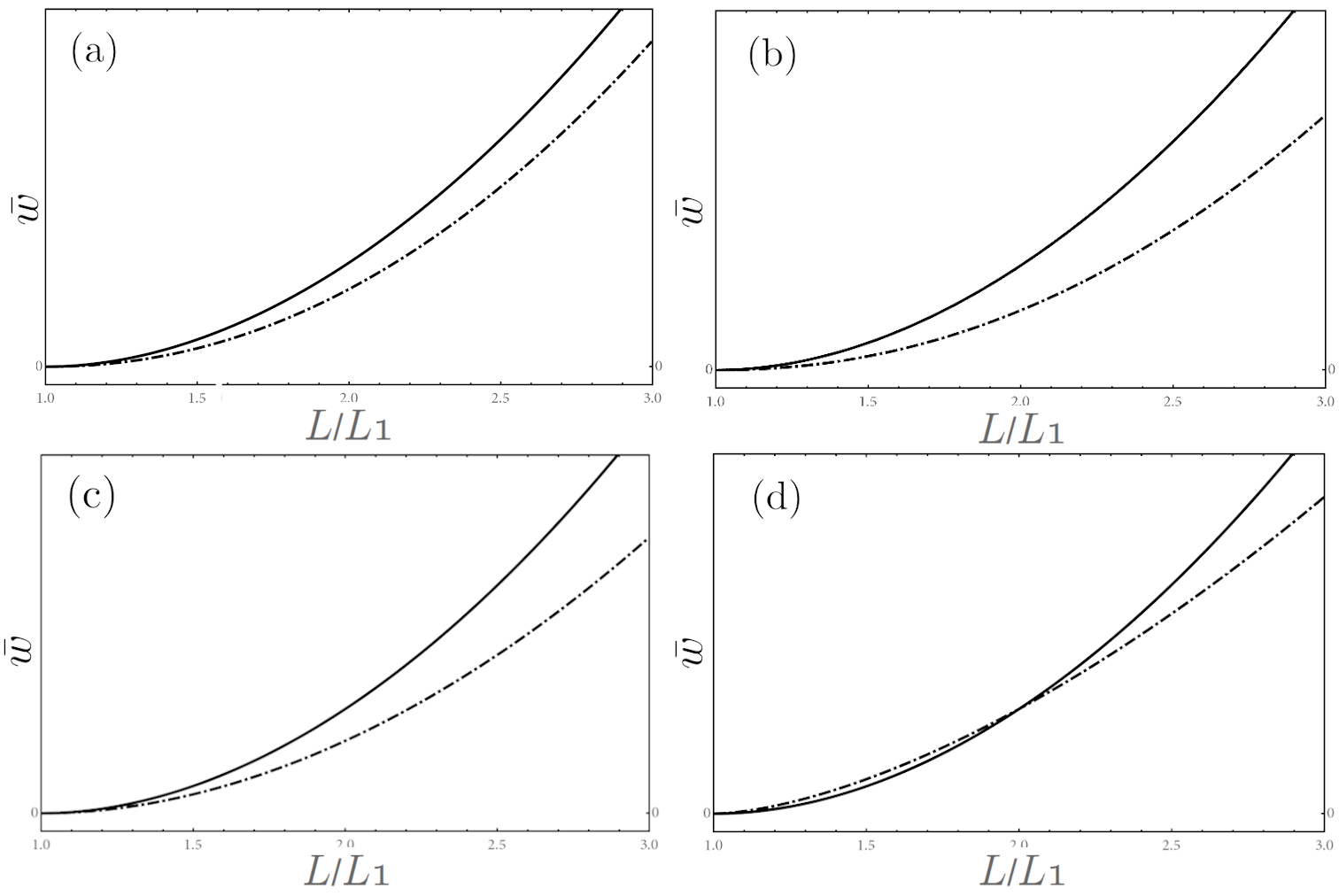}
    \caption{\textbf{The parameter dependence of the Phillips curve.}
    The solid curve is the same as in Fig.~\ref{fig:Phillips1}, while the dot-dashed curve is with the following parameters' change: 
    (a) $c$ is increased from $0.5$ to $0.9$;
    (b) $\gamma$ is decreased from $0.5$ to $0.2$; 
    (c) $B$ is increased by 20\%;
    (d) $\beta$ is increased from 0.5 to 0.6;
   }
    \label{fig:Phillips2}
\end{figure}

Now, the Philips curve with relatively small changes of the parameters $c, \gamma, B$ and $\beta$ in the manner explained above are illustrated in Fig.~\ref{fig:Phillips2} in comparison with the Phillips curve in Fig.~\ref{fig:Phillips1}.

\begin{figure}
    \centering
    \includegraphics[width=0.8\textwidth]{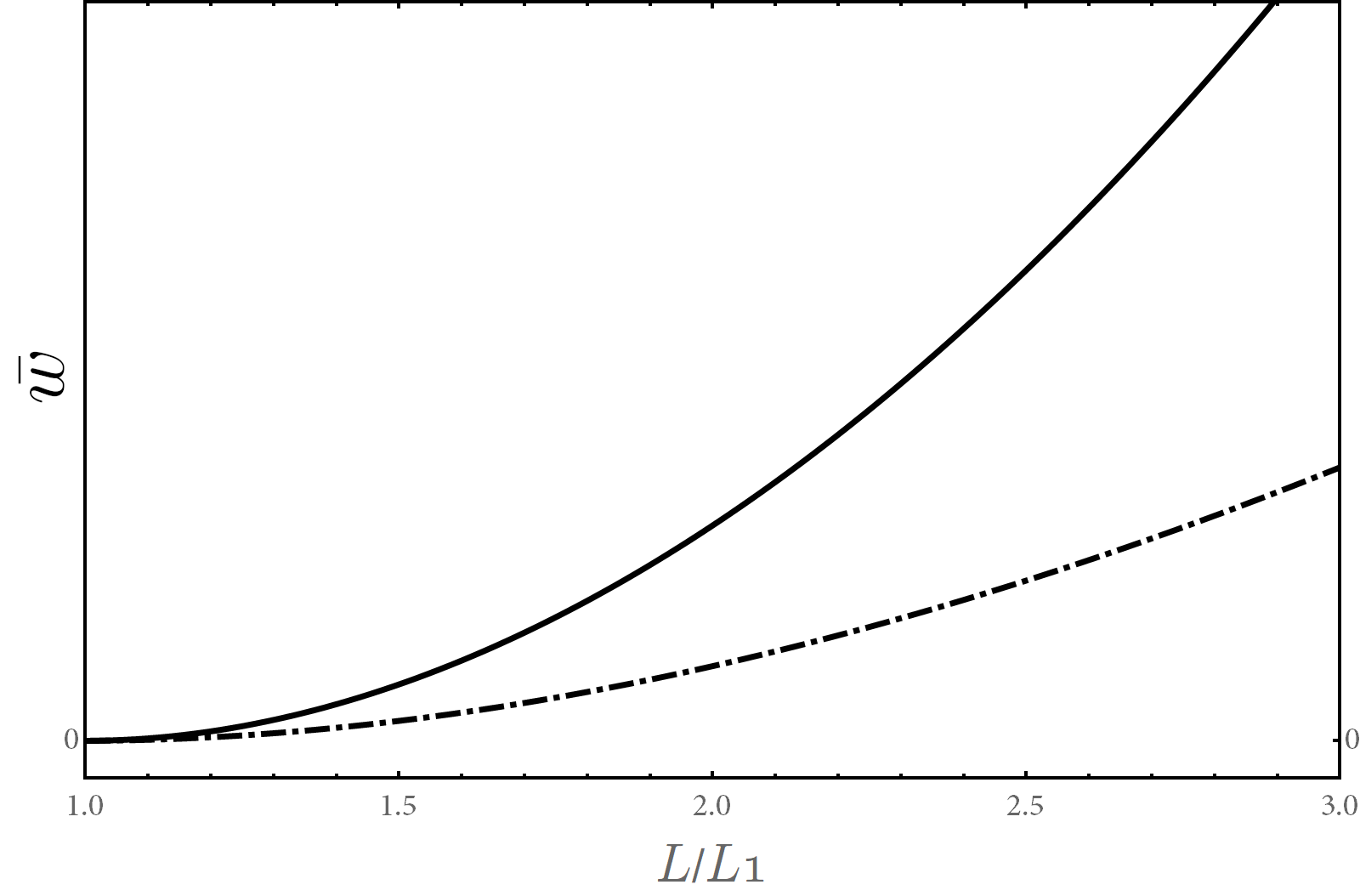}
    \caption{\textbf{Cumulative Effect of the changes of the parameters.}
    The parameters are $\alpha=c=\beta=\gamma=0.5$ for the solid curve
    and $\alpha=0.5,\ c=0.9,\ \beta=0.6,\ \gamma=0.2$ with 20\% increase in $B$ for the dot--dashed curve.
    The combined effect of small parameter changes accumulate and flatten the Phillips curve.}
    \label{fig:Phillips3}
\end{figure}

Combining all the effects of changes of four parameters $c, \gamma, B$ and $\beta$ shown
in  Fig.~\ref{fig:Phillips2}, we obtain Fig.~\ref{fig:Phillips3}, where we can clearly observe a flattening of the Phillips curve.

\clearpage

\section{Discussion}\label{sec:discussion}
We compare the results obtained in the model with actual data.
For this purpose, we focus on the first term in Eq.(\ref{eq:wbarLl}).
Take the period 2018--19 (let us denote this time-window as ``period II") when the unemployment rate was about 2.5\%, and the nominal wage growth was around 0.5\%.
The unemployment rate was about 2.5\% in the late 80's (in red in the lower panel of figure \ref{fig:PC}) to the early 90's (in green in the lower panel of figure \ref{fig:PC}, identified as ``period I") but the the nominal wage growth was around 3\%.
Given 2.5\% unemployment rate, the wage growth was lower in period II than in period I by 2,5\%:
the Philips curve had flattened.
We explore how this change is generated in our model.

\newcommand\co{c_{\rm I}}
\newcommand\ct{c_{\rm II}}
\newcommand\go{\gamma_{\rm I}}
\newcommand\gt{\gamma_{\rm II}}
\newcommand\Bo{B_{\rm I}}
\newcommand\Bt{B_{\rm II}}
\newcommand\rc{c_{\rm ratio}}
\newcommand\rg{\gamma_{\rm ratio}}
\newcommand\rB{B_{\rm ratio}}

We first discuss whether the change in the Phillips curve can be explained by varying a single parameter. For example, we want to test whether such a structural change in the curve can be entirely originated by a change in the relative productivity of the two classes of workers. 
We denote $c$ in period I and II by $c_I$ and $c_{II}$, respectively.
Let us assume that the average wage for one period is given and take one year as the time reference.
Taking period I, it increased by 3\% next year.
In period II, it increases only by 0.5\%. 
Thus the next year's wage in period II is $1.005/1.03\simeq 0.976$ times that of the next year in period I.
This difference can be achieved by increase of the value of $c$ by 2.5\%, 
since according to Eq.(\ref{eq:wbarLl}) the average wage is inversely proportional to $c$ and $c_{II}/c_{I}=1.03/1.005\simeq 1.025$.
There is no evidence suggesting that  productivity of secondary workers had increased that much.
Rather, our modes suggests that it is most plausible that the change of the Phillips curve is the result of the combined effects of simultaneous changes in the different parameters.
These changes indeed occurred in Japan between period I and period II, as discussed below.

\begin{figure}[t]
    \centering
    \includegraphics[width=0.7\textwidth]{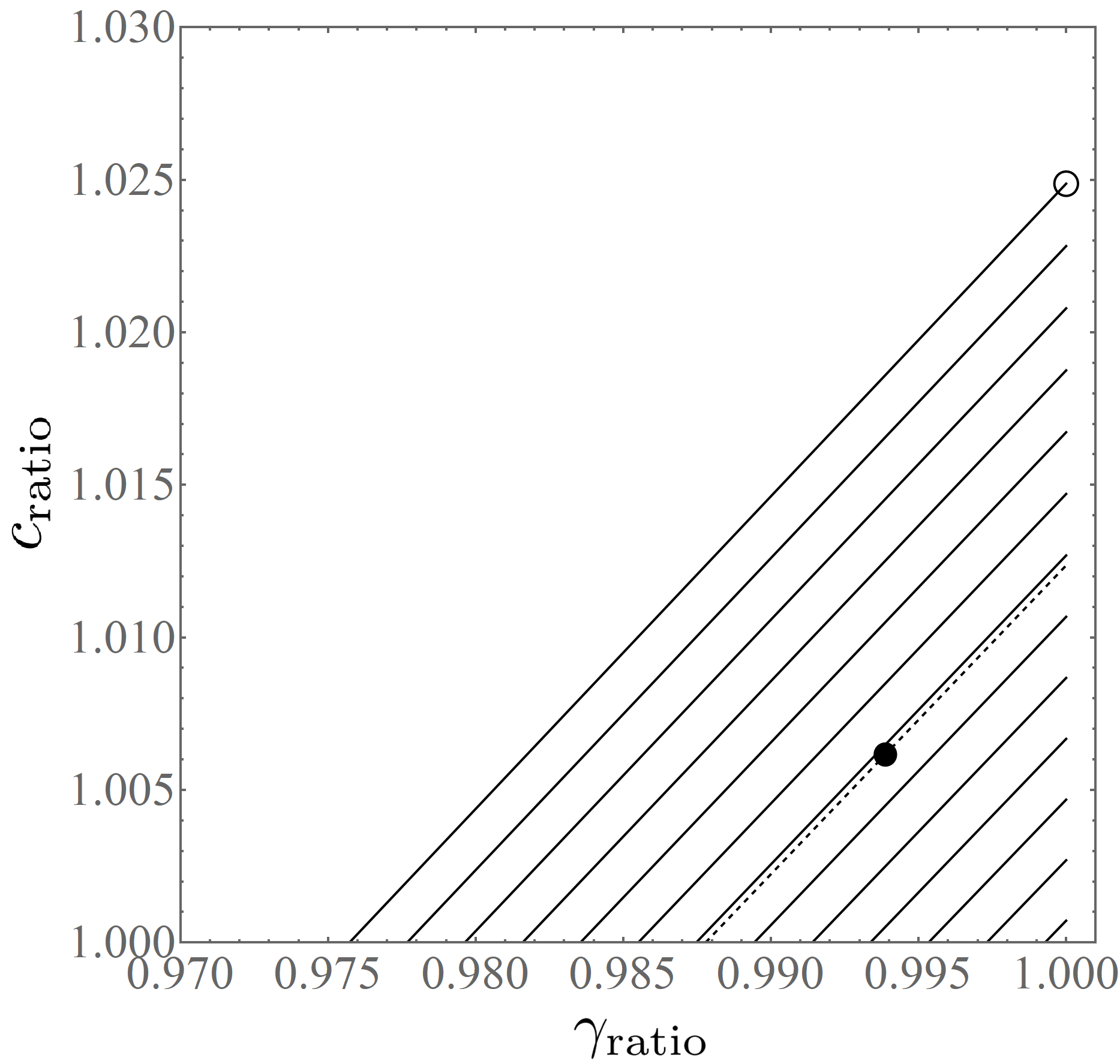}
    \caption{
    Change of parameters $\rg, \rc$ and $\rB$ that explain flattening of the Phillips curve from period I to II, given by Eq.(\ref{eq:ratios}).
   The solid lines are for $\rB=1, 1.001,1.002 \cdots$ from top to bottom.
   The circle shows the cases when only $c$ changes.
   The filled circle shows the case when $c, B$ and $\gamma$ are changed simultaneously by the similar level explained in the text. The dashed line is for the required value of change in $B$ ($\rB=1.00616$).}
    \label{fig:caseA}
\end{figure}

Now, let us discuss  how the change between those two periods is explained by simultaneous change in $c, \gamma$, and $B$, keeping $\beta$ fixed in order to focus on possible structural modifications in the labor market.
The ratios of the parameters between two periods $\rc=\ct/\co, \rg=\gt/\go, \rB=\Bt/\Bo$ 
(we denote values of parameters in each period with suffix I or II)
that explain the change of the wage growth from period I to II satisfy the following: 
\begin{equation}
    \frac{\rg}{\rc \rB^{1/\beta}}=\frac{1.005}{1.03}=0.976.
    \label{eq:ratios}
\end{equation}
If we require that the three parameters change by the same ratio, 
$\rc=\rB=1/\rg=r$ with $\beta=0.5$, we obtain $r=1.00616$,
that is, 0.62\% increase in $c$ and $B$ with 0.61\% reduction in $\gamma$.
From this we learn that even though change in each parameters is minute, it brings the wage growth down significantly, from 3\% to 0.5\%. 
The general solutions to Eq.(\ref{eq:ratios}) is illustrated in Fig.\ref{fig:caseA}.
This computational experiment suggests that the changes in the Japanese Phillips curve should be modeled as a combined effect of relatively marginal changes in the labor market.

The changes of parameters in the model which make the Phillips curve flatter are 
(1) an increase of productivity of secondary workers relative to primary workers,
(2) weaker bargaining power of primary workers,
(3) an increase of supply of secondary workers,
and
(4) an increase of wage elasticity of supply of secondary workers.
These are indeed changes which occurred in the Japanese economy over the last thirty years.

The share of secondary or irregular workers in Japan was 15--16\% during the late 1980's, but has steadily increased since then to almost 40\% in 2020
\citep{kawaguchi2013declining,Gordon2017new}.
After bubble busted at the beginning of the 1990's, Japanese firms facing unprecedented difficulties had attempted to cut labor cost by replacing highly-paid primary workers with low-wage secondary workers.
Historically, secondary workers are considered to have a lower productivity due to less training and lower attachment to the employer \citep{Fukao06,Shinada11}. 
However, in face of a dramatic increase in secondary workforce, overall productivity has been mostly stagnant and has not showed the  decline that the variation in the proportion of secondary workers would imply. Consequently, in the absence of disaggregated data, it is possible to infer that the relative productivity of secondary workers has improved, reducing the gap with primary workers' one.

While data about wage elasticity for the different categories of workers are not available, the supply of workers who are more most likely to be employed as secondary has increased, thanks in particular to two factors. First, post-war baby boomers had reached the age of retirement leaving primary jobs and entering secondary labor market. At the same time, in line with what happened in other developed economies, an increase in female participation has created additional availability of workforce, in particular for part-time employment.

The de-unionization of capitalist economies over the last decades is a well-documented phenomenon, which has recently been confirmed by micro-data analysis, as in the cited paper by \cite{stansbury2020declining}. However, Japanese unions have faced a higher pressure to restrain wage demand, in comparison to Western economies, in order to limit the loss of jobs following the bust of the double bubble in the stock and real estate markets and the lost decade.

The results have clear policy implications. The inflation target cannot be achieved without structural changes in the labor market, aimed to reverse (or at least to lessen the impact of) the changes discussed above.  In particular, the past reforms aimed to make the job market more flexible and the progressive de-unionization observed in virtually all the developed economies should be reconsidered in order to lessen the upward rigidity of wages that is affecting all the employed workforce but especially the secondary workers.

\section{Concluding Remarks}\label{sec:conclusion}
The paper presents a parsimonious model of the labor market in which the labor force is composed of temporary and permanent workers. A Phillips curve which includes the structural parameters of the labor market is analytically derived. The slope of the Phillips curve depends on the composition of the workforce, the bargaining power of workers, and the labor supply elasticity to wage of secondary workers. The solution of the model allows for a qualitative assessment of the role of a series of structural change observed in the Japanese economy on the flattening of the Phillips curve. 

In terms of policy prescriptions, the results suggest that policy maker cannot achieve target inflation without structural changes in the labor market. The present model can be included in a more comprehensive framework, in order to better highlight the feedback effects among the sluggish dynamic of wages, aggregate demand and demand for labor. Such an extension could be suitable for a more empirically grounded analysis, proving some quantitative assessment on the flattening of the Phillips curve.

\section*{Acknowledgements}
This study was supported in part by the Project ``Macro-Economy under 
\linebreak
COVID-19 influence: Data-intensive analysis and the road to recovery'' undertaken at Research Institute of Economy, Trade and Industry (RIETI), 
MEXT as Exploratory Challenges on Post-K computer (Studies of Multi-level Spatiotemporal Simulation of Socioeconomic Phenomena),
and
Grant-in-Aid for Scientific Research (KAKENHI) by JSPS Grant Numbers 17H02041 and 20H02391. 
We would like to thank Ms.\ Miyako Ozaki at Research Institute For Advancement of Living Standards for providing us with a part of the data shown in Fig.\ref{fig:PC}.

\clearpage    
\appendix
\section{Dimensional Analysis}
In this appendix, we discuss the dimensions of various parameters and variables in our model.

Dimensions play important role in various fields of natural science.
Basic dimensions in natural science are: Length, Weight, Time and Charge.
In any equation that deals with natural quantities, the dimension of the left-hand side has to be equal to the dimension on the right-hand side. For example, ``1 [in meter] = 1 [in kilogram]" does not make sense.
For this reason, we often learn a lot by simply looking at the dimensions of the constants and variables.
This is called ``dimensional analysis".

Also, dimensionless quantities play important roles in analysis: The most famous dimensionless constant is
the fine structure constant $\alpha=e^2/\hbar c=1/137.035...$ (in cgs units), 
where $e$ is the  unit of electric charge, $\hbar$ is the reduced Planck's constant and $c$ is the speed of light.
As this quantity is dimensionless, $\alpha$ has this value, regardless of whether length is  measured in meters or feet,
or whether weight is measured in kilogram or pounds, and so on.

Our analysis of the model benefits greatly by the dimensional analysis.
Let us examine dimensional properties of quantities in our model.
We denote 
the dimension of the number of workers by $\bH$, 
unit of value, like the dollar or yen, by $\bM$, 
and time by $\bT$.

First, the parameters $c, \alpha, \beta$ and $\gamma$ are dimensionless by their definitions.
Dimensions of the fundamental variables are the following:
\begin{align}
    \dim Y &= \bM\, \bT^{-1},\\
    \dim L_{1,2}&=\bH, \\
    \dim w_{1,2}&=\bH^{-1} \,\bM\, \bT^{-1},
\end{align}
as $Y$ is value created per a unit of time (yen per year, for example),
$L_{1,2}$ are number of workers,
and $w_{1,2}$ are value per person per time.
From these, we find the following dimensions of the parameters:
\begin{align}
    \dim A &= \bH^{-\alpha} \,\bM \,\bT^{-1},\\
    \dim B &= \bH^{1+\beta} \,\bM^{-\beta}\, \bT^{\beta}.
\end{align}
The former is obtained by the requirement that dimensions of the right-hand side and the left-hand side of Eq.\eqref{eq:output} matches, and the latter similarly from Eq.\eqref{eq:supply}.
From these, we find that the scaled variable $v$ (Eq.\eqref{eq:vdef})
and the parameter $g$ (Eq.\eqref{eq:gdef}) are dimensionless.
For this reason, the nonlinear equation Eq.\eqref{eq:detw2}, which plays a central role in our model but is rather complicated, is simplified to a form much simpler and easier to analyse, Eq.\eqref{eq:veq}. 

\newpage
\section{Domination of secondary workers}
Large-$g$ perturbative solution for Eq.\eqref{eq:veq} is the following:   
\begin{align} 
 v =&
 g^{-\sigma/(1+\sigma)}\biggl[\, 1
  -\frac{\sigma}{1+\sigma}\,{g^{-1/(1+\sigma)}}
    +\frac{\sigma}{2(1+\sigma)^2} \,{g^{-2/(1+\sigma)}}
    +\cdots\biggr]
    \label{eq:zs2}
\end{align}
This leads to,
\begin{align}
L&=\frac{L_1}{c}\,g^{1/(1+\sigma)}
 \left[1  +\left(c-\frac{\sigma}{1+\sigma}\right)g^{-1/(1+\sigma)}
  +\cdots\right],
\label{eq:Ltemp}
\\[5pt] 
\bar{w}
&=\left(\frac{L_1}{B}\right)^{1/\beta}
\frac{\alpha + \gamma  -\alpha\gamma}{\alpha\, c^{1/\beta}}
  g^{1/(\beta (1+\sigma))} \left[1+a_1\, g^{-1/(1+\sigma)} +\cdots\right].
\end{align}
The coefficient $a_1$ of the non-leading term in $\barw$ is a complicated function of $\alpha, \beta, c$ and $\gamma$, which is not essential for our discussion and is not given here.
Perturbative solution of Eq.\eqref{eq:Ltemp} for $g$ is the following: 
\begin{equation}
    g=\left(c\,\frac{L}{L_1}\right)^{1+\sigma}
    \left[\,1-\left(1+\sigma-\frac{\sigma}{c}\right)\frac{L_1}{L}    \cdots\right].
\end{equation}
Note that current assumption is that $L/L_1=1+L_2/L_1\gg1$.
Therefore, we find that 
\begin{align}
    \barw&=\left(\frac{L}{B}\right)^{1/\beta}
        \frac{\alpha + \gamma  -\alpha\gamma}{\alpha}
   \,\biggl[\,1+\cdots \biggr],
    \label{eq:wbarLs}
\end{align}
where the non-leading term `$\cdots$' is of order of $L_1/L$ $(\ll 1)$. 
Again we obtain a monotonically increasing Phillips curve, whose gradient is determined by $B$.

\newpage


\end{document}